\documentclass[11pt,a4paper, english,preprint,aps,nofootinbib]{revtex4}
\usepackage[english]{babel}

\usepackage[utf8]{inputenc}
\usepackage[english]{babel}
\usepackage{graphicx,color}
\usepackage{latexsym}
\usepackage{amsmath}
\usepackage{amsfonts}
\usepackage{amssymb,slashed}
\usepackage{bbold}
\usepackage[utf8]{inputenc}
\usepackage{verbatim,epigraph}
\usepackage{mathrsfs}
\usepackage{bbm}
\usepackage{bm}

\makeatletter\renewcommand{\section}{\@startsection
{section}{1}{\z@}{-3.5ex plus -1ex minus
    -.2ex}{2.3ex plus .2ex}{\bf }}

\makeatletter\renewcommand{\subsection}{\@startsection{subsection}{2}{\z@}{-3.25ex
plus -1ex minus
   -.2ex}{1.5ex plus .2ex}{\it }}
\makeatletter\renewcommand{\subsubsection}{\@startsection{subsubsection}{3}{-2.45ex}{-3.25ex
plus -1ex minus -.2ex}{1.5ex plus .2ex}{\it }}

\makeatletter \@addtoreset{equation}{section}

\newcommand{\be}{\begin{equation}}
\newcommand{\ee}{\end{equation}}
\newcommand{\bea}{\begin{array}}
\newcommand{\ea}{\end{array}}
\newcommand{\beqa}{\begin{eqnarray}}
\newcommand{\eeqa}{\end{eqnarray}}
\newcommand{\nn}{\nonumber}

\begin{document}

\vspace*{-2cm}

\title{Spontaneous Breaking of Lorentz Symmetry and Vertex Operators for Vortices}

\author{\textbf{A. P. Balachandran}\footnote{bal@phy.syr.edu} $^{a,b}$}

\author{\textbf{Se\c{c}kin K\"{u}rk\c{c}\"{u}o\v{g}lu}\footnote{kseckin@metu.edu.tr} $^{c}$}

\author{\textbf{Amilcar R. de Queiroz}\footnote{amilcarq@unb.br} $^{d}$}  

\affiliation{$^a$ Physics Department, Syracuse University, Syracuse, NY, 13244-
1130, USA}

\vspace*{1cm}

\affiliation{$^b$Institute of Mathematical Sciences, CIT Campus, Taramani, Chennai 600113, India}

\vspace*{1cm}

\affiliation{$^c$ Middle East Technical University, Department of Physics, Dumlupinar Boulevard, 06800, Ankara, Turkey}

\vspace*{1cm}

\affiliation{$^d$ Instituto de Fisica, Universidade de Brasilia, Caixa Postal 04455, 70919-970, Brasilia, DF, Brazil}

\vspace*{1.5cm}

\begin{abstract}

We first review the spontaneous Lorentz symmetry breaking in the presence of massless gauge fields and infraparticles. This result was obtained long time ago in the context of rigorous quantum field theory by \cite{Frohlich1979241,Frohlich197961} and reformulated by Balachandran and Vaidya in \cite{Bal-Sachin} using the notion of superselection sectors and direction-dependent test functions at spatial infinity for gauge transformations. Inspired by these developments and under the assumption that the spectrum of the electric charge is quantized (in units of a fundamental charge $e$), we construct a family of vertex operators which create  winding number $k$, electrically charged Abelian vortices from the vacuum (zero winding number sector) and/or shift the winding number by $k$ units. 
Vortices created by this vertex operator may be viewed both as a source and as a probe for inducing and detecting spontaneous Lorentz symmetry breaking.

We find that for rotating vortices, the vertex operator at level $k$ shifts the angular momentum of the vortex by $k \frac{{\tilde q}}{q}$, where ${\tilde q}$ is the electric charge of the quantum state of the vortex and $q$ is the charge of the vortex scalar field under the $U(1)$ gauge field. We also show that, for charged-particle-vortex composites, angular momentum eigenvalues shift by $k \frac{{\tilde q}}{q}$, ${\tilde q}$ being the electric charge of the charged-particle-vortex composite. This leads to the result that for $\frac{{\tilde q}}{q}$ half-odd integral and for odd $k$, our vertex operators flip the statistics of charged-particle-vortex composites from bosons to fermions and vice versa. For fractional values of  $\frac{{\tilde q}}{q}$, application of vertex operator on charged-particle-vortex composite leads in general to composites with anyonic statistics.
\end{abstract}

\maketitle

\section{Introduction}

The study of infrared (IR) effects in gauge field theories, especially due to its conceived intimate relation with the problem of quark confinement, has brought out several other interesting problems in quantum field theory (QFT), some of which are addressed at different levels of sophistication and some others await a proper treatment. Interesting and nontrivial effects already appear at the level of QED. The consequences of soft photons may be seen both at the level of perturbative calculations and at non-perturbative treatments. A classic example for the former is the Bloch-Nordsieck \cite{Bloch-Nordsieck} cancellation of IR divergences by soft photons, which also plays a crucial role in Schwinger's computation of the magnetic dipole moment of electrons \cite{Schwinger}. As for the latter, spontaneous breaking of Lorentz symmetry by IR photons in QED shown by Fr\"ohlich, Morchio and Strocchi \cite{Frohlich197961,Frohlich1979241} is a prominent example.

It appears that there are good physical reasons to look out for the consequences of accumulation of soft photons. Suppose that we are interested in prescribing  a well-defined initial data for a set of fields on space-time to probe their time evolution. We may consider the simpler case of a globally hyperbolic space-time $\mathbb{R}\times \Sigma$ with $\mathbb{R}$ standing for time and $\Sigma$ for a space-like hypersurface. It is possible to fix some $\Sigma_t$ as a Cauchy surface at a time $t$ where the initial conditions for the fields are given. Usually for the sake of convenience one only considers initial data supported on a compact region of the Cauchy surface $\Sigma_t$. In a more realistic situation, the accumulation of IR photons traveling from the past of the initial time $t$ must also be taken into consideration. Therefore, the initial data for the electromagnetic field, which 
are non-vanishing arbitrarily far away on $\Sigma_t$, have to be considered (Fig.1). This has consequences for the dynamics of the quantum fields, since the symmetries will be effected by the presence of these IR photons. For instance as already mentioned above, it leads to the spontaneous breaking of Lorentz symmetry in QED.

\begin{figure}
  \centering
    \includegraphics[width=0.6\textwidth, height=0.3\textheight]{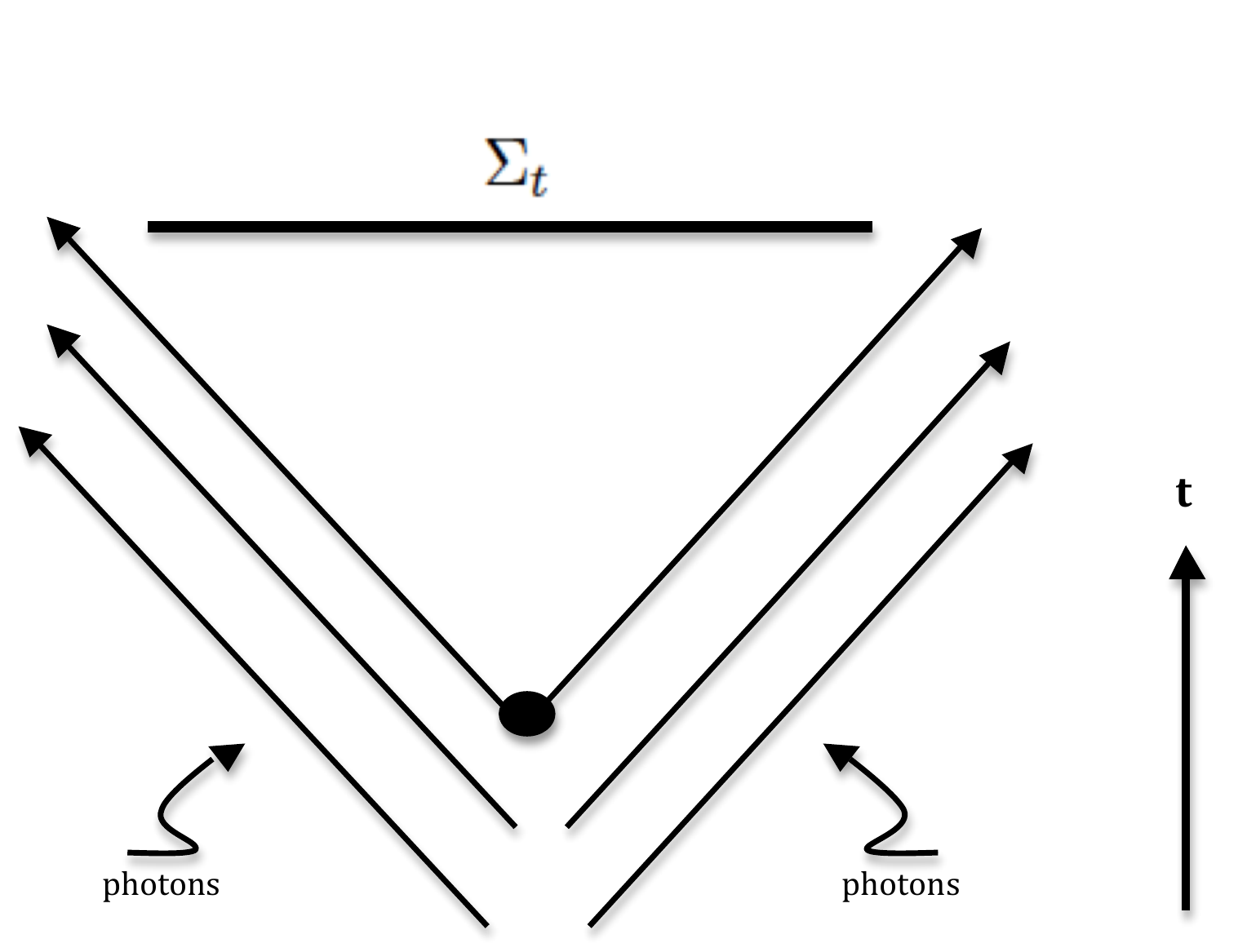}
      \caption{Clearly there are photons originating sufficiently past to $\Sigma_t$ which will arrive arbitrarily far away from any origin.}
\end{figure}

The above description of the initial data problem also appears to be relevant in the modelling of the ``night sky'' \cite{Bal-Sachin} or cosmic microwave background (CMB). CMB radiation which is supported far away from us (on scales of microscopic physics) contains IR photons coming from the distant past of the universe. It turns out that radiation with a similar physical origin is relevant for the description of the quantum states. It leads to spontaneous breakdown of Lorentz symmetry. 

Following Fr\"ohlich, Morchio and Strocchi \cite{Frohlich197961}, it is also important to stress that the spontaneous breaking of Lorentz symmetry by the accumulation of IR photons is not due to the need to introduce an IR cut-off. The argument about the resolution power of the measuring apparatus, that is, the usual text-book interpretation of the IR cut-off, cannot be evoked for the ensuing symmetry breaking. As we have argued above, such photons are associated with the setting up of a proper initial data for the dynamics of fields on space-time.

Spontaneous breaking of Lorentz symmetry in QED can be contemplated by invoking the notion of super-selection sectors \cite{Haag} and using direction-dependent test functions  at spatial infinity. Such a treatment is given by Balachandran and Vaidya in \cite{Bal-Sachin}. In this article, we will give a very brief review of some of the aspects of this work adapted for our purposes. In particular, we write down the charge operator $\widehat{Q}(\chi)$ in terms of the direction-dependent test functions $\chi$ and show that the associated eigenvalues and eigenvectors change under rotations, clearly indicating the spontaneous breaking of the Lorentz Symmetry.

In the present work we focus on the  spontaneous breaking of rotational (and hence Lorentz) symmetry in the presence of electrically charged quantum vortices in the Abelian Higgs theory in $2+1$-dimensions. Using a multivalued set of test functions $\xi$, we construct the operators
$\widehat{V}(\xi)$ by exponentiating $\widehat{Q}(\xi)$ and show that $\widehat{V}(\xi)$ are vertex operators for the creation of electrically charged vortices. Indeed, we find that these operators change the vorticity and consequently the angular momentum of dynamical vortices.  It is essential to emphasize that vortices created by $\widehat{V}(\xi)$ may be viewed as probes to detecting spontaneous Lorentz symmetry breaking. Since the latter is a phenomenon encountered in the charged sectors of QED ( as it should be clear from the previous paragraph and the discussion in \cite{Frohlich197961}) our line of reasoning is further supported by the observation that the electrically charged vortices may be viewed as sources in the context of QED leading to the spontaneous Lorentz symmetry breaking and subsequently to the construction of the vertex operators $\widehat{V}(\xi)$. In short, our results indicate a possible way in which spontaneous Lorentz symmetry breaking may be induced and probed.

Our present work is organized as follows: In section $3$ we construct a family of vertex operators which create winding number $k$ electrically charged Abelian vortices from the vacuum (zero winding number sector) and/or shift the winding number by $k$ units. Our discussion here is divided into several subsections, where rotating vortices, charged-particle-vortex composites and addition of Chern-Simons term are considered.We find that for rotating vortices the vertex operator at level $k$ shifts the angular momentum of the vortex by $k \frac{{\tilde q}}{q}$, where ${\tilde q}$ is the electric charge of the quantum state of the vortex and $q$ is the charge of the vortex scalar field under the $U(1)$ gauge field. We also show that, for charged-particle-vortex composites angular momentum eigenvalues shift by $k \frac{{\tilde q}}{q}$, ${\tilde q}$ being the electric charge of the charged-particle-vortex composite. This leads to the result that for $\frac{{\tilde q}}{q}$ half-odd integral and for odd $k$ our vertex operators flip the statistics of charged-particle-vortex composites from bosons to fermions and vice versa. For fractional values of  $\frac{{\tilde q}}{q}$, application of vertex operator on charged-particle-vortex composite leads in general to composites with anyonic statistics. 

\section{Spontaneous Lorentz Symmetry Breaking}

\subsection{Canonical Structure of Classical Electrodynamics}

We recall how the algebra of observables is constructed in the classical canonical formulation of electrodynamics. Let us start by fixing a time-slice in $3+1$-dimensional space-time. On this fixed spatial surface, the electric field $E_i$, $i=1,2,3$, is the momentum conjugate to the vector potential $A_j$. Confining our discussion to the classical theory, we can further suppose that a given charge density $J_0$ is localized in the sense that it has a compact support on the fixed spatial surface. The Gauss' law reads 
\begin{equation}
G(\Lambda)=\int d^3x~\Lambda~\left( \partial_i E_i+ J_0\right) \approx 0,
\end{equation}
where $\Lambda$ belongs to an appropriate test function space. The requirement of differentiability of $G(\Lambda)$ with respect to variations of $E_i$ determines that elements of this test function space should fulfill
\begin{equation}
\Lambda(x)\to 0 ~\textrm{ as } r\equiv |\vec{x}| \to \infty.
\end{equation}
$G(\Lambda)$ are first class constraints. They generate the $U(1)$ (Gauss law) gauge transformations:
\begin{equation}
\left\{ G(\Lambda),A_i \right\} = \partial_i \Lambda. 
\end{equation}
The associated charges are defined as
\begin{equation}
Q(\chi)=\int d^3 x~\left( -E_i \partial_i\chi + \chi J_0 \right),
\end{equation}
where the test functions $\chi$ go to a constant at spatial infinity. They are first class since they have vanishing Poisson brackets (PB) with the Gauss law
and therefore they are the observables of the theory. They generate the $U(1)$ gauge transformations with the PB's
\begin{equation}
\left\{Q(\chi_1),Q(\chi_2) \right\} = 0 \,.
\end{equation}

We emphasize that the distinction between charges $Q(\chi)$ and Gauss law constraints $G(\Lambda)$ may be cast in terms of the associated test function spaces. For  the constraints this space is composed of test functions vanishing at spatial infinity, while for the charges it is composed of functions that are constant at spatial infinity and do not necessarily vanish.

Quantization lifts $G(\Lambda)$ and $Q(\chi)$  to operators 
\be
\widehat{G}(\Lambda) \quad \mbox{and} \quad  \widehat{Q}(\chi)
\ee
acting on a suitable Hilbert space. The PB's are promoted to commutators by the usual Dirac's quantization prescription.

\subsection{Direction-Dependent Test Functions}

Accumulation of infrared photons at spatial infinity, as discussed in the introduction, gives us good physical reasons to relax the condition on the test functions for charges being constant at spatial infinity. We observe that classically 
\begin{equation}
\int d^3 x~E_i\partial_i \chi = \lim_{r\to \infty} \int d\Omega~r^2~\hat{n}_i E_i~\chi(r,\theta,\phi)  - \int d^3 x \chi \partial_i E_i \,.
\label{eq:boundaryterm}
\end{equation}
For natural choices of $E_i$ -- for instance those that fall as $\frac{1}{r^2}$ when $r\to\infty$, the surface integral in the right hand side of the above expression converges for test functions $\chi(r, \theta, \Phi)$, which has a direction-dependent limit as ${\bm r} \rightarrow \infty$:
\be
\lim_{r \rightarrow \infty} \chi(r, \theta, \phi) : = \chi(\infty, \theta, \phi) \equiv \chi(\hat n) < \infty \,.
\label{eq:dirdep1}
\ee
We henceforth accept the possibility (\ref{eq:dirdep1}). 

Spatial infinity can be viewed as a two-sphere $S^2$ and we can make the mode expansion
\begin{equation}
 \lim_{r\to\infty} \chi(r,\theta,\phi)\equiv \chi(\hat{n})=\sum \chi_{lm} Y_{lm},
\label{eq:testfunctions} 
\end{equation}
in spherical harmonics $Y_{lm}$ with $l\in \mathbb{N}$ and $-l\leq m \leq l$ as usual. Now because of the Gauss law constraint in quantum states, the effect of 
$Q(\chi)$ on quantum states depend only on $\chi(\hat n)$. Accordingly the associated ``charge" can also be mode expanded as
\begin{equation}
\label{angular-dep-charge-1}
Q(\chi)=\sum Q_{lm}Y_{lm}.
\end{equation}

Before we proceed it is worthwhile to remark that, the choice of the asymptotic behaviour $\frac{x_i}{r^3}$ for the electric field is in fact too restrictive. Suppose that we try to compute the charges $Q_{lm}$ at spatial infinity in this case. Inserting $\chi(\hat n) = \sum \chi_{lm} Y_{lm}$ in (\ref{eq:boundaryterm}), we find by orthogonality of $Y_{lm}$ that only the $Y_{00}$ component contributes to the integral, therefore at spatial infinity we seem to have only $Q_{00} \neq 0$ and the rest of the charges vanishing at infinity. However, for a more general electric field, say that due to an accelerated  charge, in general all $Q_{lm}$ survive. Then, the results of \cite{Frohlich197961, Bal-Sachin} show that the boost symmetry is in fact broken, and therefore such a boost cannot be performed in quantum theory.   

\subsection{Poincar\'e Group as an Automorphism of Algebra of Observables}

Let us turn our attention to quantum states and briefly point out how the Lorentz symmetry is spontaneously broken under the physical setting described above. For a detailed account of these results we refer the reader  to \cite{Bal-Sachin}.

A quantum state is labelled by eigenvalues of $\widehat{Q}(\chi)$, that is, the eigenvalues $Q_{lm}$ of all $\widehat{Q}_{lm}$, in addition to other relevant quantum numbers. It may be denoted as 
\begin{equation}
\label{quantum-state-1}
|\cdot; Q_{\lbrace lm \rbrace } \rangle\equiv |\cdot;Q_{00},Q_{1,-1},Q_{1,0},Q_{11},...\rangle , 
\end{equation}
where $\cdot$ represents other quantum numbers (see below) and we regard $\{lm\}$ as a collective index from now on. Observe that
\begin{align}
\widehat{G}(\Lambda) |\cdot;Q_{\lbrace lm \rbrace } \rangle & = 0, \\
\widehat{Q}_{l'm'} |\cdot; Q_{\lbrace lm \rbrace } \rangle &= \delta_{l'l}\delta_{m'm} Q_{l'm'} |\cdot; Q_{\lbrace lm \rbrace }\rangle = Q_{lm} |\cdot; Q_{\lbrace lm \rbrace }\rangle
\end{align}

A given test function $\chi(\hat{n})$ defines a superselection sector in the nomenclature of \cite{Haag}. Its choice is equivalent to fixing all the eigenvalues of $\widehat{Q}(\chi)$, that is, the $Q_{\lbrace lm \rbrace }$ in $|\cdot; Q_{\lbrace lm \rbrace } \rangle$ at spatial infinity. A given superselection sector corresponds in algebraic language to a given irreducible representation (IRR) of the net of algebra of observables. In the context of QED, $\widehat{Q}(\chi)$ with $\chi$ in a fixed space of  direction-dependent test functions at infinity gives a specific superselection sector for the algebra of observables of the theory. 

It is also important in this context to distinguish between local and global observables. Other quantum numbers of  a state vector $|\cdot;Q_{\lbrace lm \rbrace } \rangle$ are eigenvalues of some set of local observables. These local observables are defined within a region of compact support and as such they cannot change globally defined charge sectors, that is the values of $\{lm\}$; they are associated with the inner automorphisms of the algebra of observables. In contrast, all symmetries that transform states from a superselection sector to another superselection sector are said to be spontaneously broken. Such transformations can be interpreted as contained in the set of outer automorphisms. The Lorentz group (more generally the Poincar\'e group) is not a local observable, and in fact an outer automorphism acting on the quantum states and the operators ${\widehat Q}_{lm}$. As we show below for the subgroup of rotations, the action of the Lorentz group changes the superselection sector, leading to the interpretation that Lorentz group is spontaneously broken.

\subsection{Rotations}

For our purposes it is enough to restrict attention to the subgroup of rotations. For a more comprehensive discussion of the spontaneous breaking of the Poincar\'e group in QED and other related consequences due to infrared photons, we refer the reader to \cite{Frohlich1979241,Frohlich197961, Bal-Sachin}.

Rotations are implemented by the unitary irreducible representation (UIRR), $U(R)$, of the rotation group:
    \be
    {\widehat Q}_{lm}^\prime = U(R) {\widehat Q}_{lm} U(R)^{-1} =  \sum D_{m^\prime m}^l (R) {\widehat Q}_{l m^\prime} \,,
    \ee
where $D_{m^\prime m}^l(R)$ denotes the matrix elements of the rotation group element for the rotation $R$ in the (IRR) $l$. 
Under rotations the state kets transform as  
    \beqa
    U(R)  | \cdot,  Q_{\lbrace l m \rbrace } \rangle &\equiv&  | \cdot, D_{m^\prime m}^l (R) Q_{\lbrace l m^\prime \rbrace }\rangle \nn \\
    &\equiv& | \cdot, D_{m^\prime m}^{l_1} (R) Q_{l_1 m^\prime} \,, D_{m^\prime m}^{l_2} (R) Q_{l_2 m^\prime} \,, \cdots D_{m^\prime m}^{l_i} (R) Q_{l_i m^\prime}\,, \cdots \rangle \,,
    \eeqa
and sum over relevant repeated indices are implied. We note again that $\{lm\}$ is a collective index running over all possible values of $l$ and $m$. We observe that under the action of the rotation group the state kets can be rotated to another superselection sector, with a whole set of generically new charges. Hence rotational symmetry and consequently Lorentz symmetry are generically spontaneously broken. 

Note that $U(R)$ is unitary only in the direct sum of all superselection sectors it connects. Note also that if the $Q_{lm}$ in a superselection sector are all invariant under the rotation group, then the latter is not spontaneously broken in this sector.
    
In particular, if we make a rotation about the third axis, with 
    \be
    U(R) = e^{i \varphi J_3} \,, \quad D^l_{m m^\prime}(R) = \delta_{m m^\prime}e^{i m\varphi}
    \ee
    we have on the state kets 
     \be
    e^{i \varphi J_3}  | \cdot, Q_{\lbrace lm \rbrace } \rangle = | \cdot, e^{i m_1\varphi} Q_{l_1 m_1} \,, e^{i m_2\varphi} Q_{l_2 m_2} \,, \cdots e^{i m_i\varphi} Q_{l_i m_i}\,, \cdots \rangle     
    \ee
    
The above discussion may be easily adapted to $(2+1)$-dimensions. In particular, direction-dependent test functions in $(2+1)$-dimensions may be written as
    \begin{equation}
    \chi(\varphi)\equiv \lim_{r\to\infty} \chi(r,\varphi) = \sum_{n} a_n e^{in\varphi}.  
    \label{eq:testf21}
    \end{equation}
    
In this case, a finite rotation generated by $J_3$, takes the superselection sector labeled by $|\cdot; n\rangle$ to another superselection sector labeled by $|\cdot; n^\prime\rangle$, leading to the spontaneous breaking of the Lorentz group. 

As a final remark in this section, we emphasize once more that there might be a subgroup of the Poincar\'e group surviving the breaking. Which subgroup survives depends on the choice of the superselection sector: the stability group of the associated vector states contained in the Lorentz group survives unbroken.

\section{Vortices and Vertex Operators}

In this section we discuss the consequences of spontaneously broken Lorentz symmetry for $(2+1)$-dimensional vortices.
The Abelian Higgs model has the Lagrangian density 
\begin{equation}
    \label{abelian-vortex-Lag}
\mathcal L = - \frac{1}{4}F_{\mu \nu} F^{\mu \nu} + |D_0 \Phi|^2 - |D_i \Phi|^2 - \lambda (|\Phi|^2 -a^2)^2,
\end{equation}
where $\Phi$ is the complex scalar field and $D_\mu=\partial_\mu-iq A_\mu$ is the covariant derivative, with $A_\mu$ the $U(1)$ gauge potential with the field strength $F_{\mu\nu}$ and $q$ the coupling constant. Under a $U(1)$ gauge transformation the fields transform as
\be
\Phi \to e^{i \chi} \Phi \,, \quad A_\mu \to A_\mu + \frac{1}{q} \partial_\mu \chi.
\ee

Static vortex solutions in this model are characterized by the winding number of the Higgs field $\Phi$ \cite{Nielsen-Olesen}. It is standard to work in the radial gauge $A_r=0$. For rotationally symmetric configurations, a winding number $n \in {\mathbb Z}$ field profiles at spatial infinity ($r \rightarrow \infty$) as 
\begin{equation}
\Phi = a e^{in \varphi} \,, \quad r \rightarrow \infty \,.
\end{equation}
Finiteness of energy in $(2+1)$-dimensions requires that $D_\mu \Phi = 0$ as $r \rightarrow \infty$. Therefore we have,
\begin{equation}
A_\varphi = - \frac{1}{q} i \partial_\varphi \ln \Phi \quad \quad  r \rightarrow \infty \,,
\end{equation}
and the magnetic flux is proportional to the winding number:
\begin{equation}
\frac{q}{2 \pi} \oint A_\mu dx^\mu = n \,.
\end{equation}

The winding number operator can be written as 
\begin{align}
{\widehat N} &= - \frac{i}{2 \pi} \int d^2 x \, \varepsilon^{ij} \partial_i {\hat \Phi}^*  \partial_j {\hat \Phi} \nn \\
&= - \frac{i}{2 \pi} \oint_{|x| \rightarrow \infty} {\hat \Phi}^* \partial_i {\hat \Phi} dx^i \,,
\label{eq:windingnumber}
\end{align}
where ${\hat \Phi} = \frac{\Phi}{|\Phi|}$.

The charge-current density is given by
\begin{equation}
J_\mu = iq (\Phi^* D_\mu \Phi - \Phi D_\mu \Phi^*)  \,.
\end{equation}

We note that, at the classical level the charge density $J_0$ vanishes identically for a static vortex. So the electric charge (i.e. $Q_{00}$ in the notation of (\ref{angular-dep-charge-1})) vanishes for static vortices. Static vortices do not carry electric charges and have only magnetic field ${\bf B}$. In order to study the consequences of the breaking of Lorentz symmetry we need non-vanishing electric charge. This is quite natural and expected, since as emphasized in the introduction and in section 2, spontaneous breaking of Lorentz symmetry occurs only in the nonzero charged sectors of QED. Electrically charged vortices may be used to induce this symmetry breaking. There appears to be several possible ways of associating an electric charge to a vortex. In what follows we consider some of these possibilities.

\subsection{Rotating Vortices}

We can study a dynamic instead of a static vortex. We still keep the $A_0 = 0$ gauge choice. The action contains the additional electric field term $E_i^2 = (\partial_0 A_i)^2$. The equal-time commutation relations are 
\be
\lbrack A_i(x) \,, E_j(y) \rbrack = i \delta_{ij } \, \delta^{(2)}(x-y) \,, \quad 
\lbrack \Phi(x) \,, \partial_0 \Phi^*(y) \rbrack = i \delta^{(2)}(x-y) \,.
\label{eq:basiccom-1}
\ee

In addition to the conserved energy and momentum, the vortex now has angular momentum which is conserved. In other words, we have rotating vortices \cite{Manton}.

The adjoint actions of ${\widehat V}(\chi) := e^{\frac{i}{q}\widehat{Q}(\chi)}$ on the quantum fields $A_i$ and $\Phi$,
    \be
    e^{-\frac{i}{q}\widehat{Q}(\chi)} \Phi e^{\frac{i}{q}\widehat{Q}(\chi)} = e^{i\chi} \Phi \,, \quad 
    e^{-\frac{i}{q}\widehat{Q}(\chi)} A_i e^{\frac{i}{q}\widehat{Q}(\chi)} = A_i + \frac{1}{q} \partial_i \chi.
    \label{eq:adjtr}
    \ee
follow from the basic commutation relations given in (\ref{eq:basiccom-1}). Infinitesimally these read
\be
\left[\widehat{Q}(\chi),\Phi \right] = -q \chi \Phi \,, \quad  \left[\widehat{Q}(\chi),A_i \right] = i \partial_i \chi \,.
\ee

For the ensuing discussion, we assume that the spectrum of the conserved non-vanishing electric charge 
\begin{equation}
{\widehat Q}_e := \frac{1}{e} \int  d^2 x \, J_0 \,
\end{equation}
is quantized in units of a fundamental charge $e$. That is,
\begin{equation}
\textrm{Spec}~ {\widehat Q}_e = {\mathbb Z} \,.
\label{eq:chargequan}
\end{equation}
On quantum states ${\widehat Q}_e := \frac{1}{e} {\widehat Q}(\chi)$ where $\lim_{r \rightarrow \infty} \chi ({\hat n}) = 1$. It also worth remarking that ${\widehat Q}_e - \frac{1}{e} {\widehat Q}_{00}$ is a Gauss law constraint.

Suppose that we consider $\widehat{Q}(\xi)$, where $\xi (\varphi) =  k \varphi \,, k \in {\mathbb Z}$, which is clearly {\it not} in the space of functions (\ref{eq:testf21}), since it is multivalued on $S^1$. Consider the operator 
\beqa
    \label{vertex-op-def-1}
 \widehat{V}(\xi) \equiv e^{\frac{i}{q} \widehat{Q}(\xi)} &=& e^{\frac{i}{q}\int d^2 x ~ (- \partial_i  \xi E^i + \xi J_0 )} \nn \\
 &=& e^{\frac{i k}{q}\int \frac{1}{r} dr d \varphi ~(-E_\varphi + \varphi J_0 )} \,,
\eeqa
where the second line is written in polar coordinates in the coordinate basis with the metric components $h_{rr} = 1 \,,  h_{\varphi \varphi} = r^2 \,, h_{r \varphi} = 0$.

It appears to represent a singular gauge transformation, but we show below that it generates a well-defined finite gauge transformation. Hence local observables commute with it and its eigenvalues also serve to define superselection sectors. We will see, however that rotations do not commute with it. Hence rotations, and consequently Lorentz transformations get spontaneously broken.

In what follows we demonstrate that it is in fact a vertex operator for vortices. 
Due to (\ref{eq:chargequan}) it is a well-defined operator with the property
\beqa
{\widehat V}(\xi + 2 \pi k ) &=& e^{\frac{i k}{q} \int \frac{1}{r} dr d \varphi \,  (- E_\varphi + \varphi J_0) + \frac{2 \pi i k}{q} \int d^2 x \, J_0 } \nn \\
&=& e^{i 2 \pi k \frac{m e}{q}} {\widehat V}(\xi)
\label{eq:consistency}
\eeqa
The phase on the right hand side classifies the possible statistics associated with these operators, as will be discussed in the subsequent section. 

For future reference we note that
\beqa
{\widehat V}(\xi + 2 \pi k ) &=& {\widehat V}(\xi) \quad \mbox{for} \quad k m \frac{e}{q} \in {\mathbb Z} \nn \\
{\widehat V}(\xi + 2 \pi k ) &=& - {\widehat V}(\xi) \quad \mbox{for} \quad k m \frac{e}{q} \quad \mbox {half odd integer} \,.
\eeqa
The operator (\ref{vertex-op-def-1}) is indeed gauge invariant since
\begin{equation}
 \left[\widehat{G}(\Lambda),\widehat{Q}(\xi)\right]\approx 0.
 \label{eq:gaugeinv}
\end{equation}

We now proceed to interpret ${\widehat V}(\xi) = e^{\frac{i}{q} {\widehat Q}(\xi)}$ as a vertex operator shifting the winding number of the vortex by $k$ units. 

The action of the operator (\ref{vertex-op-def-1}) on $ \hat \Phi$ and $D_\mu \hat \Phi$ reads
    \begin{align}
    {\widehat V}(\xi)^{-1} \hat \Phi {\widehat V}(\xi) &= e^{i k \varphi} ~  \hat \Phi, \nn \\
    {\widehat V}(\xi)^{-1}\left( D_\mu  \hat \Phi \right) {\widehat V}(\xi) &= e^{i k \varphi}~D_\mu  \hat \Phi.
    \end{align}
Using the above formula we obtain
\begin{align}
{\widehat V}(\xi)^{-1} \left ( \partial_\mu {\hat \Phi} \right) {\widehat V}(\xi) &= {\widehat V}(\xi)^{-1} D_\mu {\hat \Phi} {\widehat V}(\xi) + i q{\widehat V}(\xi)^{-1} A_\mu {\hat \Phi} {\widehat V}(\xi) \nn \\
&= e^{i k \varphi} D_\mu {\hat \Phi}+ i q {\widehat V}(\xi)^{-1} A_\mu {\widehat V}(\xi) {\widehat V}(\xi)^{-1}   {\hat \Phi} {\widehat V}(\xi) \, \nn \\ 
&= e^{i k \varphi} D_\mu {\hat \Phi} + i q (A_\mu + \frac{1}{q} k \partial_\mu \varphi) e^{i k \varphi} {\hat \Phi} \, \nn \\
&= e^{i k \varphi} (\partial_\mu {\hat \Phi} + i  k (\partial_\mu \varphi ){\hat \Phi}) \,.
\end{align}

From (\ref{eq:windingnumber}) and (\ref{eq:adjtr}) we find
\begin{align} 
{\widehat V}(\xi)^{-1} {\widehat N} {\widehat V}(\xi) &= {\widehat N} + \frac{1}{2 \pi} \oint_{|x| \rightarrow \infty} k \partial_i \varphi dx^i \, \nn \\
&= {\widehat N} + \frac{k }{2 \pi} \int_0^{2 \pi} \partial_\varphi \varphi d \varphi \, \nn \\
&= {\widehat N} + k \,.
\end{align}
which together with (\ref{eq:gaugeinv}) clearly indicate that ${\widehat V} (\xi)$ is a gauge invariant vertex operator shifting the winding number by $k$ units.

The above result may also be expressed as
\begin{equation}
\lbrack {\widehat V}(\xi) \,, {\widehat N} \rbrack = - k {\widehat V}(\xi) \,.
\end{equation}

Denoting the quantum state of the vortex with winding number $n$ as $| n \,, \cdot \rangle$, we have ${\widehat N} | n \,, \cdot \rangle = n | n \,, \cdot \rangle$ and also 
\begin{equation}
\lbrack {\widehat V}(\xi) \,, {\widehat N} \rbrack | n \,, \cdot \rangle = - k {\widehat V}(\xi)  | n \,, \cdot \rangle
\end{equation}
from which we obtain
\begin{equation}
{\widehat N} ({\widehat V}(\xi)  | n \,, \cdot \rangle ) = (n + k ) {\widehat V}(\xi)  | n \,, \cdot \rangle 
\end{equation}
indicating that
\begin{equation}
{\widehat V}(\xi)  | n \,, \cdot \rangle \equiv  | n + k \,, \cdot \rangle.
\end{equation}

It is also worthwhile to remark that in the literature there are the so-called `` 't Hooft loop operators", they are given as $ T (C^\prime) = e^{- i \int_{C^\prime} d^3 x \, \alpha_i E_i}$, for a suitable one-form $\alpha= \alpha_i dx^i$ \cite{tHooft, Nair}, and act as vortex creation operators in a manner similar to the operators introduced in this work. We are not going to make any detailed comparison of these two operators at present, however, it is readily observed that in our case for ${\widehat V}(\xi)$ the role of $\alpha_i$  is played by $ \partial_\varphi \xi$ which is closed in the integer cohomology ${\cal H}(S^1 \,, {\mathbb Z})$, whereas for the `` 't Hooft loop operators" this not the case; $\alpha$ is not closed $d \alpha \neq 0$. 

\subsection{Shift of Angular Momentum}

Let us consider a rotating vortex described by the quantum states $|\tilde{q},j\rangle$ satisfying
\be
\widehat{Q}_e |\tilde{q} \,, j \rangle = \frac{\tilde{q}}{e} |\tilde{q} \,, j \rangle \,, \quad J_3 |\tilde{q} \,, j \rangle  = j |\tilde{q} \,,  j \rangle.
\label{eq:eigenv1}
\ee
We are ignoring other possible quantum numbers irrelevant for our discussion.

Under a spatial rotation by $\theta$, that is $\varphi \rightarrow \varphi + \theta$, we have
    \beqa
    {\widehat V}(\xi) \rightarrow e^{i \theta J_3} {\widehat V}(\xi) e^{- i \theta J_3} &=& e^{\frac{i k}{q} \int \frac{1}{r} dr d \varphi (- E_{\varphi + \theta})}  e^{\frac{i}{q} \theta k \int d^2 x J_0} \, \nn \\ 
    &=& {\widehat V}(\xi) e^{i \frac{e}{q} \theta k \widehat{Q}_e} \,,
    \label{eq:shift1}
    \eeqa
where in passing from the first line line to the second line, we have made use of a change of variables, to absorb the shift in the argument of $E_\varphi$, since $d (\varphi + \theta) = d \varphi$. 
     
It is now easy to compute 
    \begin{equation}
    {\widehat V}(\xi) |\tilde{q} \,, j \rangle = e^{\frac{i}{q} {\widehat Q}(\xi)}  |\tilde{q} \,, j \rangle.
    \end{equation}
We have from (\ref{eq:eigenv1}) and (\ref{eq:shift1}),
    \be
     e^{i \theta J_3} {\widehat V}(\xi) |\tilde{q} \,, j \rangle  
     = e^{ i \theta (k \frac{\tilde{q}}{q} + j)} {\widehat V}(\xi)  |\tilde{q} \,, j \rangle \,,
    \label{J3-shifting-1}
    \ee
    \begin{equation}
    {\widehat V}(\xi) |\tilde{q} \,, j \rangle = |\tilde{q} \,, j+ k \frac{\tilde{q}}{q} \rangle \,.
    \end{equation}
We conclude that the action of the vertex operator on the quantum state $|\tilde{q} \,, j \rangle$ shifts the angular momentum of the state by $k \frac{\tilde{q}}{q}$.

\subsection{Charged-Particle-Vortex Composites}

Suppose now that a charged particle say with charge $\tilde{q}$ is orbiting the vortex. According to Wilczek's results \cite{Wilczek}, the angular momentum of the charged particle orbiting around a vortex shifts from integer value $j$ (in the absence of the vortex) by an amount $- \frac{\tilde{q}}{q} n$, $n$ being the winding number of the vortex. 

We can denote the quantum state of this composite system as $|\tilde{q}, J \rangle$ with
\be
\widehat{Q}_e |\tilde{q} \,, J \rangle = \frac{\tilde{q}}{e} |\tilde{q} \,, J \rangle \,, \quad  J_3 |\tilde{q} \,,  J \rangle  = J |\tilde{q} \,,  J \rangle \,,
\ee
where
\begin{equation}
J = j -  \frac{\tilde{q}}{q} n \,, \quad j \in {\mathbb Z} \,.
\label{eq:specj1}
\end{equation}
Proceeding in a similar manner as before, we find
\beqa
e^{i \theta J_3} {\widehat V}(\xi) |\tilde{q} \,, J \rangle  &=& e^{i \theta J_3} {\widehat V}(\xi) e^{- i \theta J_3}  e^{ i \theta J_3} | \tilde{q} \,, J \rangle \nn \\
&=& {\widehat V}(\xi) e^{\frac{i}{q} \theta k  \int d^2 x J_0}  e^{ i \theta J}  |\tilde{q} \,, J \rangle \nn \\
&=& {\widehat V}(\xi) e^{i \theta  k \frac{\tilde{q}}{q}}  e^{ i \theta J}  |\tilde{q} \,, J \rangle \nn \\
&=& e^{ i \theta (k \frac{\tilde{q}}{q} + J)} {\widehat V}(\xi)  |\tilde{q} \,, J \rangle \,.
\eeqa
so that we have
\begin{equation}
{\widehat V}(\xi) |\tilde{q} \,, J \rangle = |\tilde{q} \,, J + k \frac{\tilde{q}}{q} \rangle \,.
\end{equation}
The spectrum of angular momentum after the application of the vertex operator becomes
\beqa
J_{new} &=&  j -  \frac{\tilde{q}}{q} (n - k) \,, \nn \\
&=&  J + \frac{\tilde{q}}{q} k \,.
\label{eq:specj2}
\eeqa

This result has very interesting consequences. We first note that $\tilde{q}$ and $q$ come in integer multiples of the fundamental charge $e$. Let us suppose then that the ratio $\frac{\tilde{q}}{q}$ is a half-odd integer:
    \begin{equation}
    \frac{\tilde{q}}{q}  = \frac{2 \ell + 1}{2} \,, \quad \ell \in {\mathbb Z} \,.
    \end{equation}
In this case we first observe that ${\widehat V}(\xi) $ is anti-periodic for any integer $k$, that is,
    \begin{equation}
    {\widehat V}(\xi + 2 \pi k ) = - {\widehat V} (\xi) \,.
    \end{equation}

Then the spectrum of $J_{new}$ is
    \begin{equation}
	\mbox{Spec} \left ( J_{new} \right ) \equiv
	\left \lbrace
	\begin{array}{llll}
	\mbox{integer} & \mbox{for} & n \,, k \in \mbox{even integers} \\
	\mbox{integer} & \mbox{for} & n \,, k \in \mbox{odd integers} \\
	\mbox{half-odd integer} & \mbox{for} & n \in \mbox{odd integers} & k \in \mbox{even integers} \\
	\mbox{half-odd integer} & \mbox{for} & n \in \mbox{even integers} & k \in \mbox{odd integers} \\
	\end{array}
	\right.
    \end{equation}

We infer from (\ref{eq:specj1}) and (\ref{eq:specj2}),  where $n \in {\mathbb Z}$, that for any odd integer $k$, the action of  ${\widehat V}(\xi)$ takes a half-integral angular momentum state to an integral angular momentum state, and an integral angular momentum state to a half-integral angular momentum state. For even $k$, no such shift occurs, that is, integral and half-integral angular momentum states remain as integral and half-integral after the action of $V(\xi)$ for even $k$.
	
Suppose that  we interchange two identical charged particle-vortex composites. General results obtained by \cite{Wilczek} state that the statistics is normal (bosonic or fermionic depending on the statistics of the charged particle) for integral angular momentum eigenvalues and normal statistics is reversed for half-odd integral angular momentum eigenvalues. In the intermediate cases, the composites are neither bosons nor fermions; they are then anyons. 
	
From our results, we see that  for $\tilde{q}/q$ half-odd integral and odd $k$, the action of ${\widehat V} (\xi)$ on charged particle-vortex composites changes their angular momentum from integer values to half-odd integer values or vice versa. Therefore ${\widehat V} (\xi)$ flips the statistics of two such identical composites from fermions to bosons or vice versa. For example, if initially the two identical charged particle-vortex composites are fermions, then they behave as bosons after the application of ${\widehat V} (\xi)$ for odd $k$ on each composite, the winding number $n$ being integral.

We may also obtain anyonic charged particle-vortex composites starting from a bosonic or a fermionic identical pair. This can happen when $\tilde{q}/q$ is a generic rational number since
\begin{equation}
	{\widehat V}(\xi + 2 \pi k ) =  e^{i 2 \pi k  \frac{{\tilde q}}{q}} {\widehat V}(\xi) \,.
\end{equation}
	
\subsection{Adding the Chern Simons term}

Another interesting possibility is to switch on a Chern-Simons (CS) term. The details of the vortex models and the solutions are somewhat different in this case. There are several models with vortex solutions. Regardless of these details, let us briefly summarize the essential features in this case. 

Adding the Chern-Simons term
\be
L_{CS} = \frac{\kappa}{2} \varepsilon^{\mu \nu \rho} A_\mu \partial_\nu A_\rho \,,
\ee
the Gauss law constraint becomes
\begin{equation}
\int d^2 x ~ \Lambda \left(\partial_i E_i - \kappa B  + J_0 \right) \approx 0 \,, \quad B = \frac{1}{2} \varepsilon^{ij} F_{ij} \,.
\end{equation}
where $\Lambda$ are test functions vanishing at spatial infinity. 

As shown in \cite{Manton, Khare}, when the Chern-Simons term $L_{CS}$ is present, there is a non-vanishing electric charge $Q = \int d^2 x J_0$ even for static vortices, which turns out to be proportional to the magnetic flux: 
\be
Q = \kappa \int d^2 x B = \kappa \frac{2 \pi n}{q} \,.
\ee
Since this result incorporates the quantization of the electric charge into the theory, it provides an additional incentive in the present context. 

The charges are
\begin{equation}
Q(\chi) =   \int d^2 x~ \left( - E_i \partial_i \chi + \kappa \varepsilon^{ij} \partial_i \chi A_j + \chi J_0\right) \,,
\end{equation}
and they are gauge invariant.

For the operator ${\widehat V}(\xi)$ we find
\be
{\widehat V}(\xi + 2 \pi k ) =  e^{i 2 \pi k  \frac{2 \pi \kappa}{q^2} n} {\widehat V}(\xi) \,.
\ee 
We therefore, observe that ${\widehat V}(\xi)$ leads in general to anyonic charged particle-vortex composites when $L_{CS}$ is present. Let us also note that for a non-abelian theory it is well-known that the Chern-Simons level $\kappa$ is quantized as $\kappa = \frac{q^2 r}{4 \pi}$ where $r$ is an integer. In such a case appropriate non-Abelian generalisations of ${\widehat V} (\xi)$ could be periodic or anti-periodic depending on $k r n $ being an even integer or an odd integer. 

\newpage

{\bf \large Acknowledgements}

\vskip 1em

APB was supported by DOE under grant number DE-FG02-85ER40231 and by the Institute of Mathematical Sciences, Chennai. S. K is supported by TUBiTAK under project No. 110T738 and TUBA-GEBIP program of The Turkish Academy of Sciences. ARQ is supported by CNPq under process number 307760/2009-0.

\vskip 1em

\providecommand{\href}[2]{#2}\begingroup\raggedright

\endgroup

\end{document}